\documentclass[conference]{IEEEtran}
\IEEEoverridecommandlockouts
\usepackage{cite}
\usepackage{amsmath,amssymb,amsfonts}
\usepackage{algorithmic}
\usepackage{graphicx}
\usepackage{textcomp}
\usepackage{xcolor}
\usepackage{hyperref}
\usepackage{booktabs}
\def\BibTeX{{\rm B\kern-.05em{\sc i\kern-.025em b}\kern-.08em
    T\kern-.1667em\lower.7ex\hbox{E}\kern-.125emX}}

\makeatletter \newcommand{\linebreakand}{ \end{@IEEEauthorhalign} \hfill\mbox{}\par \mbox{}\hfill\begin{@IEEEauthorhalign} } \makeatother

\begin{document}

\title{Atari Games Challenge: A Pilot Study on Multimodal Player Experience Assessment}

\author{
\IEEEauthorblockN{1\textsuperscript{st} Oleg Jarma Montoya}
\IEEEauthorblockA{\textit{Software Quality Research Group} \\
\textit{IT University of Copenhagen}\\
Copenhagen, Denmark \\
olejar@itu.dk \\
0009-0008-1036-5069}
\and
\IEEEauthorblockN{2\textsuperscript{nd} Erica Manca}
\IEEEauthorblockA{\textit{Center for Digital Play - BrAIn Lab} \\
\textit{IT University of Copenhagen}\\
Copenhagen, Denmark \\
manc@itu.dk}
\linebreakand
\IEEEauthorblockN{3\textsuperscript{rd} Thomas Vase Schultz Volden}
\IEEEauthorblockA{\textit{Center for Digital Play - BrAIn Lab} \\
\textit{IT University of Copenhagen}\\
Copenhagen, Denmark \\
thvo@itu.dk \\
0009-0009-2177-5482}
\and
\IEEEauthorblockN{4\textsuperscript{th} Paolo Burelli}
\IEEEauthorblockA{\textit{Center for Digital Play - BrAIn Lab} \\
\textit{IT University of Copenhagen}\\
Copenhagen, Denmark \\
pabu@itu.dk \\
0000-0003-2804-9028}
}

\maketitle

\begin{abstract}
We present a pilot study on the collection and synchronisation of multimodal data for player experience investigation. We collected game telemetry, self-reported surveys, biometrics, and cued-retrospective think-aloud (C-RTA) data from 19 participants playing three Atari 2600 games. The study then uses the data to investigate difficulty in PX, showcasing a protocol for future multimodal research.

The dataset obtained from the experiment, which is publicly available, shows potential as a rich, transformative source that can be used to investigate dynamic difficulty adjustment algorithms, game balancing strategies or broader explorations of games user research. The study findings suggest that the experimental approach holds strong potential for generalisation in future player experience studies.
\end{abstract}

\begin{IEEEkeywords}
Human-computer interaction, Data collection, Experimental design, Difficulty
\end{IEEEkeywords}

\section{Introduction}


Player experience (PX) can be considered one of the most complex examples of human-computer interaction (HCI), as the interaction between the user, or \textit{player}, and the game is not limited, nor focused, to usability and intuitiveness, but rather aims at enabling the player to have a unique \textit{aesthetic} experience~\cite{niedenthal_what_2009}, that rarely could be achieved outside of the magic circle~\cite{gillin_homo_1951}.
Emotions, sensations, and affections are only some of the layers that build on the complexity of Player Experience, and, although extensive research has provided methods to assess the subjective state of a player~\cite{medlock_overview_2018}, there is still a large untapped potential in psychophysiological sources~\cite{burelli_playing_2024}.
To fully grasp the multifaceted nature of PX, it is essential to draw on multiple, complementary sources of data ---each valuable on its own, but most powerful when used in combination with the others.

Although there are PX data sets, such as GAMEEMO~\cite{alakus_database_2020}, they are limited to one source, in this case, physiology, and do not include any in-depth gameplay information. Other datasets, such as AGAIN~\cite{melhart_arousal_2022}, include extensive information about the in-game experience and self-reported labels, but no physiological data.

To fill the gap of well-rounded multimodal PX datasets, we elaborated an experimental protocol with the intention of multimodal data collection of PX, gathering from sources like game telemetry and surveys, psychophysiology and qualitative methods.
In this paper, we present the pilot study \textit{Atari Games Challenge}, we elaborate on the protocol used and the data collected, and as a means to demonstrate its usability, we showcase an investigation using the resulting dataset on how \textit{difficulty} presents itself in player experience.

\section{Background and Related Studies}
\subsection{Datasets in Player Experience}
Several existing datasets provide important precedents for multimodal analysis of player experience. For example, Rashed et al.~\cite{rashed2025descriptor} present a multimodal player engagement dataset that synchronises EEG, eye tracking, heart rate, gameplay frames, controller inputs, and webcam footage, enabling engagement classification across diverse game scenarios. Fanourakis et al.~\cite{fanourakis2025amucs} introduce AMuCS, an affective multimodal Counter-Strike dataset that combines physiological signals (e.g., ECG, EDA, respiration) and behavioural signals (e.g., facial expressions) to model emotional responses during competitive first-person shooter gameplay.

Other game-based affective datasets have primarily focused on emotion recognition from neurophysiological signals. The GAMEEMO database~\cite{alakus_database_2020}, for instance, provides EEG data from 28 participants playing four computer games designed to elicit boredom, calmness, fear, and humour, along with Self-Assessment Manikin (SAM) ratings of arousal and valence. Related work on EEG emotion recognition in videogame play further explores pattern recognition approaches to classify affective states from brain activity during interactive sessions~\cite{rodriguez2015eeg}.

The Atari Games Challenge Pilot Study also connects to a broader body of research on affective games and multimodal affect classification. Surveys of affective gaming outline how modalities such as eye tracking, speech, EEG, and GSR can be combined to detect arousal and valence and to enable affect-aware game mechanics~\cite{yannakakis2023affective}. Hamdy's multimodal classification system~\cite{hamdy2018affective}, for example, demonstrates that eye tracking can outperform speech in affect detection, and that decision-level fusion further improves performance in games designed to elicit controlled affective states. By including both eye tracking and pupil data alongside EEG and fine-grained gameplay telemetry, the Atari Games Challenge Pilot Study offers a complementary testbed for multimodal fusion approaches focused specifically on difficulty-related experiences.

Finally, the use of the Player Experience Inventory (PXI)~\cite{haider_minipxi_2022} situates this work within the tradition of validated player experience measurement tools. The PXI and associated resources, such as PXI Bench, operationalise player experience as a multidimensional construct spanning functional aspects (e.g., challenge, ease of control, clarity of goals) and psychosocial aspects (e.g., mastery, autonomy, meaning). The availability of benchmark PXI data for diverse games further supports comparative analyses of player experience across titles and genres. By integrating PXI-based self-reports with high-resolution physiological and gameplay data, the Atari Games Challenge Pilot Study contributes a dataset that is well aligned with established measurement frameworks, while enabling fine-grained investigation of how objective and subjective difficulty jointly shape player experience in classic arcade-style games.

\subsection{Difficulty as a multimodal presence}
The concept of difficulty in digital games has been a subject of interest for game designers and academics alike, for almost as long as digital games themselves. It has been defined in several ways and used as groundwork for multiple fields of game-related research. One of the most well-known and studied is Dynamic Difficulty Adjustment (DDA), a field heavily inspired by the theory of ``flow''~\cite{csikszentmihalyi_flow_1990}. Given a player's skill, a DDA algorithm attempts to automatically and subtly change variables in the game so the experience playing it is viewed as ``not too easy, not too hard''~\cite{zohaib_dynamic_2018}. Most attempts at DDA algorithms try to maximise a player's level of engagement by looking at their performance ---a limiting view of how difficulty works that has been criticised~\cite{guo_rethinking_2024}. Recent research has split difficulty into two types~\cite{guo_exploratory_2025}:
\begin{itemize}
\item \emph{Objective Game Difficulty (OGD)}, defined as the player's probability of failing specific in-game tasks or actions, measured via performance metrics~\cite{li_adaptive_2014}.\item \emph{Subjective Game Difficulty (SGD)}, defined as the reflection of a player's own views on their performance, which can only be collected via self-reports~\cite{denisova_measuring_2020}.
\end{itemize}
Very few studies have analysed the relation between these two sides of difficulty, and even fewer have their user data easily available.



\section{The Study Protocol}
Our study responds to the research gaps previously mentioned by designing an experimental protocol for gathering multimodal data, contributing to the evolving landscape of PX. It was supported by an internally developed software platform that handles game emulation, data acquisition, and survey delivery within a single pipeline.

\subsection{Software Framework}
The \textit{Psycho-Atari} framework was designed to help with the implementation of PX experiments related to games research and similar~\cite{atari-games-challenge-pilot-git}. It uses the \textit{Arcade Learning Environment} (ALE)~\cite{bellemare_arcade_2013,c_machado_revisiting_2018} for the emulation of \textit{Atari 2600} games, and takes advantage of its logging capabilities to recover the RAM state each frame. It also allows for the construction of survey scripts and the recollection of responses, with supported questions like multiple choice, free answer and Likert scale. Finally, it connects to physiological hardware via API calls and also records its signals in sync with the gameplay data. More advanced configurations include the reconstruction of gameplay recordings and the detection of custom events.

\subsection{Atari Games Challenge}
A recruiting campaign was conducted at the IT University of Copenhagen, the experiment was advertised through the university's campaign monitors, the weekly newsletter addressed to the whole corpus of students, printed fliers distributed across the buildings of the university, and direct recruitment via word of mouth.
A total of $19$ students took part in the pilot, of which $11$ identified as male, $7$ as female, and $1$ as non-binary. Participants' ages ranged from 19 to 32 (median 24). All participants expressed a varied level of interest regarding digital games, but none had previous experiences with the three games involved with the experiment.
Participation in this experiment was voluntary, but only candidates who did not suffer from any health condition which affected their sensitivity to flashing lights, nor any severe respiratory, cardiovascular or cerebrovascular diseases were accepted. Furthermore, participants would be rewarded with a voucher for a drink in the university coffee shop, of the value of $30$\textit{DKK} (around $4$\textit{€}), alongside free water bottles if they so desired.

\subsubsection{Setup} \label{sec:setup} The participant is introduced to the environment, answers a demographics survey and is subsequently equipped with the \textit{g.tec g.HIAMP} 64-channel EEG headset, which will record the participant's brain activity during the whole duration of the experiment.

\subsubsection{Gameplay session} \label{sec:gameplay} 
The gameplay part of the experiment takes place in a controlled environment, a sound-proofed and electrically isolated cabin, in which the participant sits in front of a 23-inch display, an \textit{Atari 2600} controller and a numeric pad. The collection of quantitative data for the analysis between OGD and SGD takes place here.


The gameplay portion of the experiment consists of one tutorial session and three gameplay sessions, each comprising three trials. A trial is defined as a single two-minute playthrough of one game at one difficulty level. Three \textit{Atari 2600} games were selected for the study, each with three pre-configured difficulty levels. Within each session, the participant plays all three games in a randomised order, with a difficulty level drawn at random and without replacement. Over three sessions, each participant completes nine trials in total, randomly covering all game--difficulty combinations. Prior to the first session, the players go through a tutorial session in which they play each game at its easiest difficulty level to familiarise themselves with the controls and mechanics.

During the whole experiment, the participants are observed by the researchers through a live-view camera with no recording functionality enabled, while two types of physiological information are recorded: electroencephalography (EEG), through the \textit{g.tec g.HIAMP} EEG headset, and eye movement, with the \textit{Gazepoint GP3 HD} remote tracker. Both of these sources are also recorded and synchronised with the rest of the data via \textit{Lab Streaming Layer} (LSL)~\cite{kothe_lab_2024}.

The games played during the experiment were chosen to represent different forms of challenge that players would be faced with. Denisova et al. \cite{denisova_measuring_2020} recognise four types of challenge in games:

\begin{itemize}
    \item \textit{Performative}: Challenge that requires accuracy and reflexes. It relates to general physical aptitude.
    \item \textit{Cognitive}: Challenge that requires effective planning, memorization, and multitasking.
    \item \textit{Decision-Making}: Challenge that requires the ability to make complicated decisions and adapt to the results.
    \item \textit{Emotional}: Challenge which requires strong resolve in front of the game's characters and story.
\end{itemize}
Out of these, we selected the first 3 as the most relevant to the experiment.

As for choosing a candidate for each challenge, we were limited to the selection of games supported by the ALE. After a deep review of the library, and taking into consideration mechanics and possible metrics to use, the following games were chosen:

\emph{Performative Challenge - \textit{Boxing}}~\cite{bob_whitehead_boxing_1980}\\
A simulation of the physical sport. The player takes control of a boxer and faces a computer-controlled rival. Both the player and the rival can move in 8 directions around the screen and throw punches. Each effective punch to the opposing entity awards several points depending on what type of hit it was: long jabs award one point, while close-up punches give 2. The game runs for 2 minutes, and at the end of this, the entity with the most points is declared the winner. Its pre-configured difficulties change either the player's or the rival's speed. The performance metric is the number of points the player obtains

\emph{Cognitive Challenge - \textit{Word Zapper}}~\cite{henry_will_iv_word_1982}\\
A shooter mixed with a spelling game. The player controls a ship that can move in 8 directions and shoot left, right, and up. A word is shown at the bottom of the screen for the first 4 seconds of the game; after that, it'll disappear. The player then has to shoot letters from a rolling alphabet at the top of the screen to spell the word. During this, meteorites of distinct shapes appear from both sides of the screen. If the player gets hit, different negative effects occur. As this game requires memorization and multitasking capability to spell correctly and avoid objects, it fits well as a representative of cognitive challenge. Its pre-configured difficulties affect which meteorites can appear, along with its spawn rate. Given the amount of time for each trial, a participant will only have time to complete one word unless they have great skill and understanding of the game. This makes the performance a Boolean variable of success/failure.

\emph{Decision-Making Challenge - \textit{Turmoil}}~\cite{sirius_software_turmoil_1982}\\
A rail-style Shooter. The player takes control of a ship with the goal of shooting aliens to score points. All aliens have different shapes and interact differently with the player; some can only be shot from the back, or not at all. Initially, the player is placed at the centre of the screen and can only move up and down between rails. However, a prize will randomly appear on a rail, allowing the player to move to the left or right to take it and go back to the center. The player cannot shoot while outside of the center, leaving themselves vulnerable if an alien appears. Also, if the player takes too long to pick the prize, it'll turn into an alien. Knowing when to shoot, positioning, and deciding how to tackle certain aliens are good examples of Decision-making challenges. The pre-configured difficulties deal with alien spawn rates. In the case of performance, the raw score cannot be used, as the difficulty configuration has an effect on the enemy spawn rate, which heavily contributes to the score. It was decided to calculate a "Score/Death Ratio", as we thought this would both mitigate the inflation caused by the difficulty and also better represent when a player does well.

\begin{align}
    R = \frac{\text{Score}}{\text{Deaths}+1}
\end{align}

\begin{figure}[!ht]
    \centering
    \begin{tabular}{ccc}
        \includegraphics[width=0.3\linewidth]{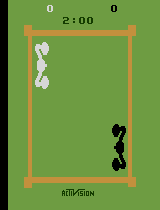} & \includegraphics[width=0.3\linewidth]{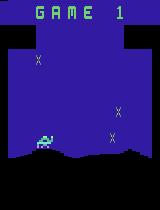} & \includegraphics[width=0.3\linewidth]{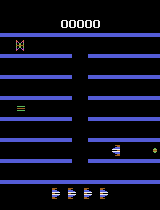}
    \end{tabular}
    \caption{Display of the selected games. From left to right: \textit{Boxing}, \textit{Word Zapper}, \textit{Turmoil}}
    \label{game_displays}
\end{figure}

Figure \ref{game_displays} displays how the games will look to the participants.

\subsubsection{Self-reports}
After every single trial is finished, a participant has to answer a survey about their experience. The \textit{miniPXI}~\cite{haider_minipxi_2022} scale was used as the post-game survey. It consists of 10 questions which each represent a different construct of Player Experience. The Questions are in a Likert Scale from -3 to 3. We are particularly interested in the constructs of ``Immersion'', ``Mastery'' and ``Challenge'', as these are the ones more closely related to difficulty; the whole survey needs to be delivered for it to be valid.

\subsubsection{C-RTA session} \label{CRTA}
After the gameplay section is over, a short break occurs, then the C-RTA session takes place.
The researchers will show the participant a selection of videos from their previous gameplay sessions.
These videos are watched twice each: the first time, without the need to communicate to the researchers, so that they can focus on remembering what they were feeling in that moment of gameplay, especially keeping in mind moments in which they felt confused, curious, or in which their expectations were unmet by the game; the second time, they would perform a think-aloud about what they remembered about the experience, following what is presented on the screen.
The videos are ordered by game, and the participant is informed whenever the trial changes.


The code necessary to replicate the experimental procedure is available on GitHub at \href{https://github.com/itubrainlab/atari-games-challenge-pilot}{https://github.com/itubrainlab/atari-games-challenge-pilot}.

\section{The Dataset}
The dataset collected during this study is publicly available on Zenodo\footnote{\url{https://doi.org/10.5281/zenodo.17699259}} as the Atari Games Challenge Pilot Dataset~\cite{it-university-of-copenhagen-2025-17699260}.
It includes recordings from 19 participants playing three Atari 2600 games at three distinct difficulty levels, resulting in 19 tutorial sessions and 57 gameplay sessions each containing three gameplay trials. During each session, physiological and behavioural measures were continuously recorded. Immediately after each session, participants completed standardised affective self-reports assessing perceived experience and challenge.

\subsection{Data Organisation and Format}
The dataset adheres to the Brain Imaging Data Structure (BIDS) specification for multimodal human neuroscience data~\cite{gorgolewski_brain_2016}. The root directory contains global metadata (\textit{dataset\_description.json}), participant demographics (\textit{participants.tsv}), and documentation (\textit{README.md}). Individual participant folders \(\textit{sub-$<$subject\_id$>$}\) include both physiological and behavioural data files organised by modality.

\subsubsection{Modalities and Measurements}
\begin{itemize}

\item \textit{EEG}: 63 channels recorded according to the 10--20 electrode placement system, referenced to the left earlobe, sampled at 256 Hz.
\item \textit{Eye Tracking}: 16-channel gaze and fixation data (left/right coordinates, validity flags) sampled at 150 Hz.
\item \textit{Pupilometry}: 21-channel recordings of pupil diameter, position, and validity indicators, sampled at 150 Hz.
\item \textit{Gameplay Telemetry}: 128-channel binary representations of game state variables (RAM bits), sampled at 30 Hz.
\item \textit{Self-Report Surveys}: 10 variables covering player experience constructs derived from the PXI (Player Experience Inventory) instrument, completed after each gameplay trial.

\end{itemize}

\subsection{Data Access and Encoding}
All data are provided in open, interoperable formats---EDF, TSV, and JSON---encoded in ASCII/UTF-8 and compliant with BIDS conventions. Each recording includes well-defined metadata and event markers supporting alignment across modalities.

\section{Data Analysis and Results}
We proceed to show the analysis done using the dataset.

\subsection{Analytical Framework}

A central goal of this study is to move beyond simple performance thresholds and instead examine \textit{how} and \textit{why} difficulty affects player experience. To do so, we draw on \textit{causal inference}, a family of methods concerned not merely with whether two variables are associated, but with whether a change in one variable produces a change in another. In an experimental setting like ours, where difficulty level is actively manipulated, causal inference provides a principled framework for estimating the effect of that manipulation on player experience outcomes, accounting for the structure of the data-generating process rather than treating all associations as equivalent.

A key question in this context is whether the effect of difficulty on player experience is \textit{direct}, meaning if it's felt by the player independently of how they actually perform or whether it operates \textit{through} performance: a harder game causes the player to score worse, and it is that change in performance that shapes how they feel. This distinction motivates the use of \textit{mediation analysis}, an extension of causal inference that decomposes the total effect of a cause into these two pathways. Formally, a variable $M$ is said to mediate the relationship between a treatment $X$ and an outcome $Y$ if part of the effect of $X$ on $Y$ operates via $X \rightarrow M \rightarrow Y$, rather than through the direct path $X \rightarrow Y$ alone. In our case, $X$ is difficulty, $M$ is in-game performance, and $Y$ is a self-reported player experience construct.

\subsection{Three-Model Workflow}

To operationalise mediation analysis, we follow a three-stage modelling workflow applied to each outcome of interest:

\subsubsection{Total effect model} The outcome (e.g., perceived challenge) is regressed on difficulty and session order alone.  This estimates the overall effect of the experimental manipulation on the outcome, without distinguishing how that effect arises.

\subsubsection{Mediator model} Performance is regressed on difficulty and session order. This tests whether the manipulation actually moves the proposed mediator.

\subsubsection{Direct effect model} The outcome is regressed on difficulty, session order, \textit{and} performance simultaneously. By comparing the coefficients on difficulty in this model against those in the total effect model, we can determine how much of the total effect is carried by performance (the indirect path) and how much remains independent (the direct path).

This workflow is applied separately for each game, and for three player experience constructs: \textit{immersion}, \textit{mastery}, and \textit{challenge}.

\subsection{Research Questions}

The modelling workflow is organised around a conditional sequence of research questions. We first ask whether the experimental manipulation has any discernible effect on the player experience outcomes. Only if that is the case does it become meaningful to ask whether performance mediates that effect.

\medskip
\subsubsection{Stage 1 --- Effect of manipulation on experience (Total effect model)}
\begin{itemize}
    \item \textit{$RQ_1$}: Does difficulty or session order affect a player's perceived \textit{immersion}?
    \item \textit{$RQ_2$}: Does difficulty or session order affect a player's perceived \textit{mastery}?
    \item \textit{$RQ_3$}: Does difficulty or session order affect a player's perceived \textit{challenge}?
\end{itemize}

\medskip
\subsubsection{Stage 2 --- Effect of manipulation on performance (Mediator model)} For any outcome where a directional effect is
found in Stage 1:
\begin{itemize}
    \item\textit{$RQ_4$}: Does difficulty or session order affect a player's in-game performance?
\end{itemize}

\medskip
\subsubsection{Stage 3 --- Mediation (Direct effect model)} If both the outcome and the mediator respond to the manipulation:
\begin{itemize}
    \item \textit{$RQ_5$}: Is the effect of difficulty and session order on player experience \textit{mediated} by in-game performance? In other words, does accounting for performance reduce or eliminate the effects observed in Stage 1?
\end{itemize}

\subsection{Bayesian Regression}

To estimate the models described above, we adopt a \textit{Bayesian} approach to regression. Unlike classical frequentist models, which produce a single point estimate and a $p$-value, Bayesian regression treats every model parameter as a probability distribution. The result of fitting a Bayesian model is a \textit{posterior distribution} over each coefficient, showing a full picture of which values are more or less consistent with the observed data, given our prior assumptions. This makes Bayesian regression particularly well-suited to causal inference: rather than looking for significance, we aim to understand the size and direction of the probability distribution, which is a more informative question when the goal is to understand mechanisms rather than simply detect associations.

A further practical advantage is that Bayesian models naturally accommodate small samples. Rather than relying on asymptotic approximations, posterior distributions are estimated via simulation, specifically through \textit{Markov Chain Monte Carlo} (MCMC) sampling, which remains valid regardless of sample size~\cite{oganisian_practical_2021, brooks_handbook_2011}. We use the PyMC probabilistic framework for both the model definitions and MCMC simulation~\cite{salvatier_probabilistic_2016}.

Prior distributions are set to be weakly informative, as we do not have any previous knowledge or expectations about the data. However, prior predictive checks were conducted to verify that the priors implied plausible distributions of miniPXI responses before any data were observed. Table \ref{tab:model-config} shows a full list of the configurations, including the choice of prior.

\begin{table}[h]
\centering

\label{tab:model-config}
\renewcommand{\arraystretch}{1.2}
\begin{tabular}{ccccc}
\toprule
\textbf{Stage} & \textbf{Outcome} & \textbf{Regression} &
\textbf{Coeff. prior} \\
\midrule
1, 3 & miniPXI  & Ordinal & $\mathcal{N}(0,\;3)$ \\
\midrule
2 & Score  & Poisson &
    $\mathcal{N}(0,\;3)$ on $\log$ scale \\
\midrule
2 & Success & Logistic &
    $\mathcal{N}(0,\;3)$ on logit scale \\

\bottomrule

\end{tabular}
\caption{Model configurations per workflow stage. All treatment variables follow a Dirichlet prior. Stages 1 and 3 share the same configuration; they differ only in whether performance is included as a predictor.}
\end{table}

Results are reported via the \textit{highest density interval} (HDI) at the 94\% level. The HDI is the shortest interval that contains 94\% of the posterior probability mass for a given parameter, or the range of values the data most support. We interpret the HDI as follows:

\begin{itemize}
    \item When the HDI lies entirely above zero, the posterior is consistent with a positive effect: higher values of the predictor are associated with higher values of the outcome across 94\% of the plausible parameter space.

    \item When the HDI lies entirely below zero, the posterior is consistent with a negative effect.

    \item When the HDI spans zero, the data suggest no effect exists.
\end{itemize}

\subsection{Results}

\begin{figure}[!ht]
    \centering
    \includegraphics[width=\linewidth]{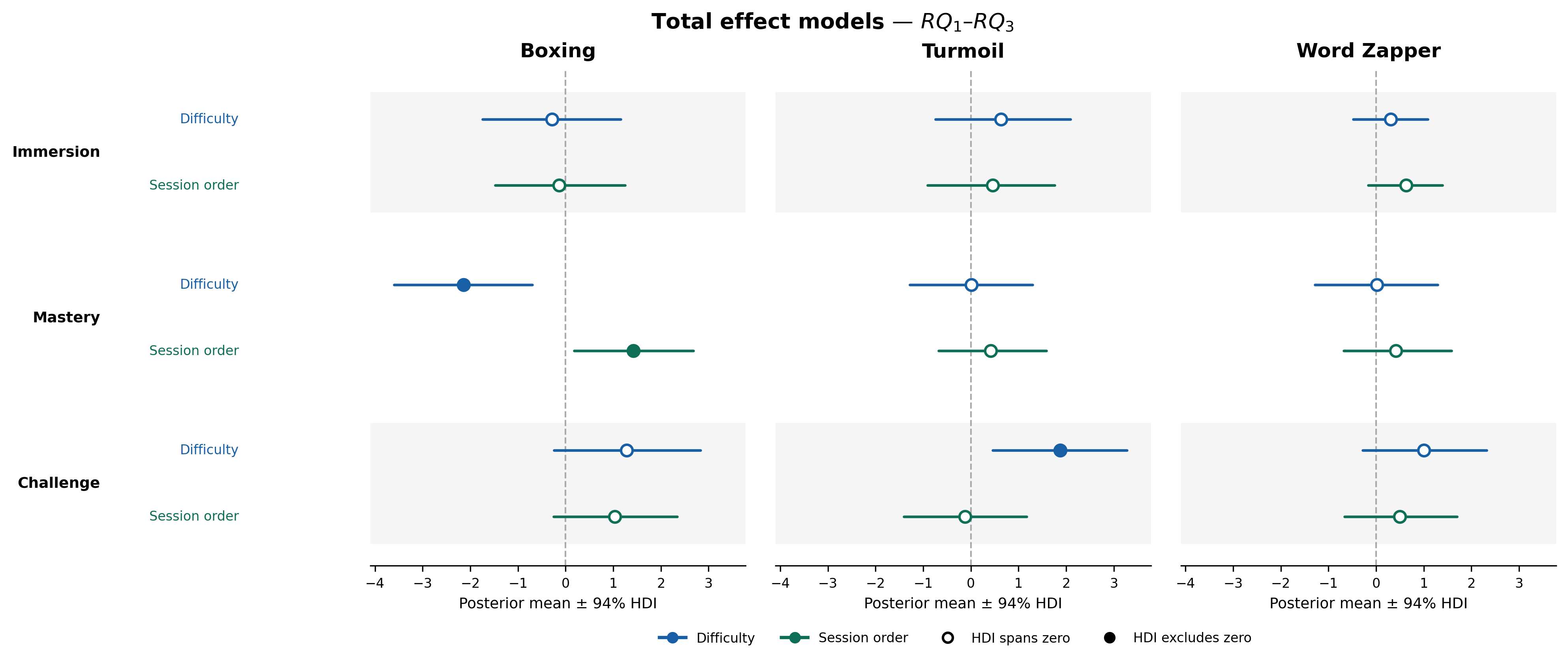}
    \caption{Total effect model posteriors for $RQ_1$--$RQ_3$, across all three games. Each row shows the posterior mean (dot) and 94\% HDI (bar) for one predictor--outcome pair. Filled markers indicate HDIs that exclude zero; open markers indicate HDIs that span zero. Blue = difficulty; teal = session order.}
    \label{fig:forest_total}
\end{figure}

\begin{figure}[!ht]
    \centering
    \includegraphics[width=\linewidth]{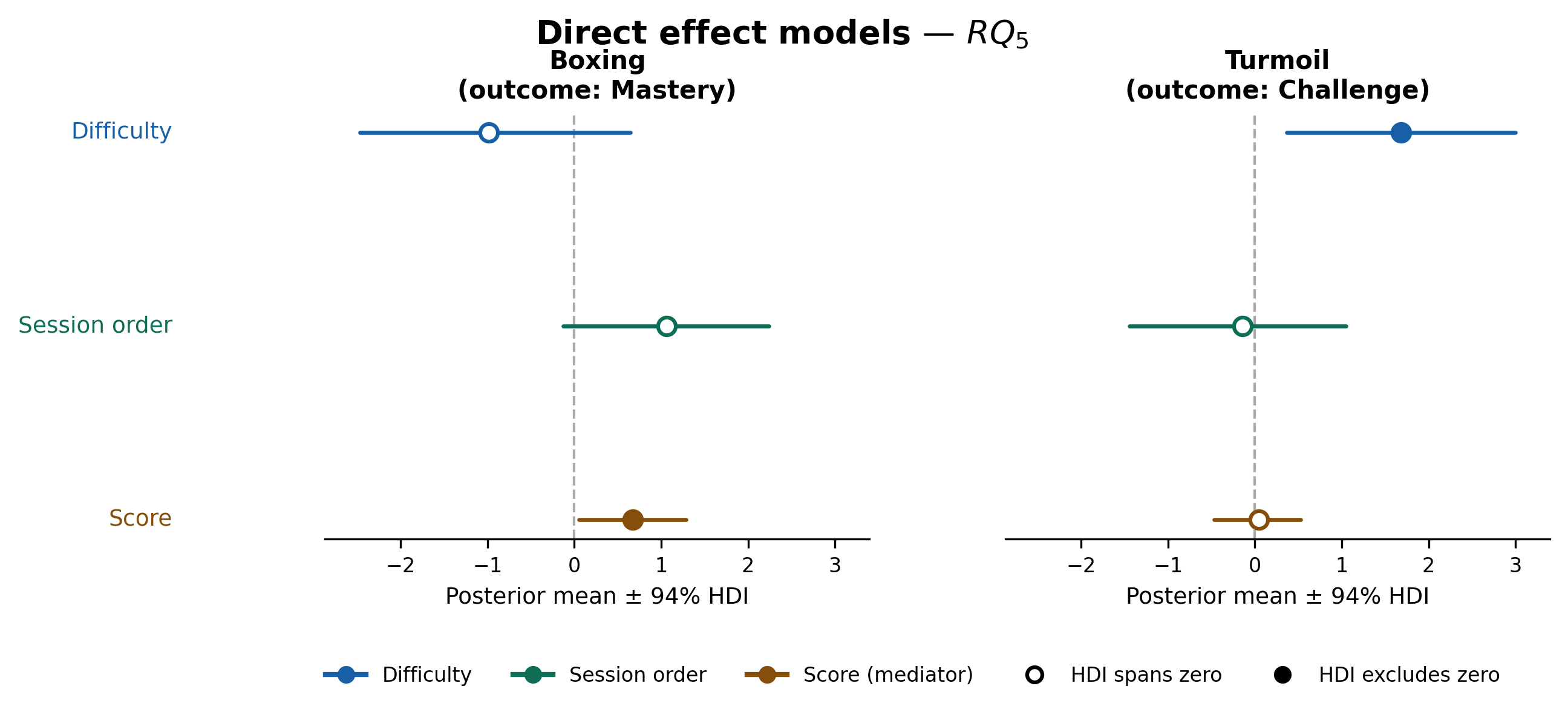}
    \caption{Direct effect model posteriors for $RQ_5$, fitted for Boxing (outcome: Mastery) and Turmoil (outcome: Challenge). Amber = score (mediator); blue = difficulty; teal = session order. Filled and open markers follow the same convention as Figure~\ref{fig:forest_total}.}
    \label{fig:forest_direct}
\end{figure}

Figure~\ref{fig:forest_total} shows the posterior means and 94\% HDIs for all total effect models ($RQ_1$--$RQ_3$) across the three games. Figure~\ref{fig:forest_direct} shows the corresponding direct effect models ($RQ_5$) for Boxing and Turmoil. In both figures, filled markers indicate that the HDI excludes zero; open markers indicate that the HDI spans zero.

\subsubsection{Boxing}

Inspecting the Boxing panel in Figure~\ref{fig:forest_total}, the posteriors for Immersion and Challenge show no clear directional pattern for either predictor as both HDIs span zero and are wide relative to their means, indicating that the data provide little information about the direction of any effect on these constructs. Mastery tells a different story: the difficulty HDI lies entirely below zero (mean $= -2.14$, 94\% HDI: $[-3.60,\,-0.70]$), while the session order HDI lies entirely above zero (mean $= 1.43$, 94\% HDI: $[0.18,\,2.68]$). The two predictors pull in opposite directions, meaning greater difficulty is associated with lower perceived mastery, while progression through sessions is associated with higher perceived mastery. This pattern is qualitatively consistent with flow theory.

For $RQ_4$, the mediator model shows that both treatment variables are associated with higher scores: each difficulty increase corresponds to approximately a 3\% increase in score, while each additional session corresponds to approximately a 25\% increase, suggesting that cross-game practice is the stronger driver of performance.

Turning to the Boxing panel in Figure~\ref{fig:forest_direct}, the picture shifts substantially once performance is included as a predictor ($RQ_5$). The difficulty and session order HDIs both move to span zero, while the score HDI lies above zero (mean $= 0.67$, 94\% HDI: $[0.05,\,1.28]$). This pattern is consistent with full mediation: the associations that difficulty and session order had with Mastery in the total effect model appear to operate through their effect on performance. For Boxing, in-game score may therefore carry most of the information relevant to a player's sense of mastery.

\subsubsection{Turmoil}

In the Turmoil panel of Figure~\ref{fig:forest_total}, the posteriors for Immersion and Mastery closely resemble those of Boxing, with both predictors showing HDIs spanning zero with no clear directional pattern. The standout result is for Challenge: the difficulty HDI is entirely above zero (mean $= 1.87$, 94\% HDI: $[0.46,\,3.28]$), while session order shows no comparable pattern. Notably, this effect is in the opposite direction to Boxing's mastery result and does not straightforwardly align with flow theory. A plausible explanation is that lower difficulty levels were not perceived as meaningfully demanding, with the game only feeling appropriately challenging at harder settings. This highlights the potential balancing issues rather than a sensitivity to the manipulation.

The mediator model for $RQ_4$ shows difficulty as the stronger predictor of score ratio, with approximately a 16\% increase per difficulty level; session order contributes approximately 10\%. Given that Turmoil's difficulty is tied to enemy spawn rate, higher scores under harder conditions are partially expected, though the fact that the custom performance metric did not fully absorb this effect suggests participants also incurred additional losses.

Looking at Figure~\ref{fig:forest_direct}, the direct effect model for Challenge shows that the difficulty posterior is largely unchanged from the total effect model (mean $= 1.68$, 94\% HDI: $[0.36,\,3.00]$), while the score HDI spans zero and is centred near zero (mean $= 0.04$). Unlike Boxing, there is no evidence of mediation here: performance does not appear to carry the difficulty effect on perceived challenge. A plausible interpretation is that participants were not attending to the score as a feedback signal, judging difficulty by enemy density alone, which would decouple performance from subjective experience.

\subsubsection{Word Zapper}
The Word Zapper panel of Figure ~\ref{fig:forest_total} is very uniform, with all six HDIs spanning zero, with the posterior means being quite small, and the intervals a lot narrower than the other two games. The data are uninformative about the direction of any treatment effect on Immersion, Mastery, or Challenge. As the conditions for proceeding to $RQ_4$ and $RQ_5$ are not met, the mediator and direct effect models are not fitted for this game.

The most plausible explanation for this pattern is regarding the game's mechanical complexity. Word Zapper requires simultaneous memorisation, spatial navigation, and target selection; participants may not have developed sufficient command of the core mechanics within the two-minute session window for the difficulty manipulation to register as a meaningful variation in experience. In this sense, the null result is itself informative: it suggests that mechanic comprehension is a prerequisite for difficulty to shape subjective experience, and that neither OGD nor SGD is a useful lever for a game players do not yet understand.

\subsection{Comparison with C-RTA}
The C-RTA session results were later revised in an effort to further confirm the results found in the Bayesian models.

In \textit{Boxing}, it was noted that participants tried their best to maximise their scores, either trying to land precise hits or finding strategies that would earn them the most points in less time. This aligns with the mediation model, where the score absorbs the effect of both difficulty and practice.  In \textit{Turmoil}, participants often ignored the score, focusing on shooting; they judged difficulty by spawn density. Finally, in \textit{Word Zapper}, although participants generally understood the goal of the game, they frequently misinterpreted or overlooked key mechanics: this lack of understanding helps explain why neither difficulty nor trial had any discernible impact on any of the \textit{miniPXI} constructs of interest.

Taken together, the Bayesian models and the C-RTA commentary illustrate that OGD and SGD often interact, but in genre-specific ways, highlighting the limitations of treating difficulty solely as a scalar target for optimisation. Instead, more nuanced, multi-perspective approaches are needed when designing DDA systems:

\begin{itemize}
    \item \textit{Performative} contexts may benefit from using performance metrics directly to inform adjustments: when in-game score fully mediates, a DDA system could track real-time scoring trends and adapt opponent speed or reaction windows to keep players in a target competency band.

    \item \textit{Decision-making} contexts might require calibrating enemy spawn rates in response to player feedback about challenge, rather than exclusively basing changes on kill/death ratios---since players' subjective sense of challenge did not align with their performance.

    \item \textit{Cognitive} contexts demand greater consideration of tutorials or mechanic clarity before meaningful causal relations can be established: if players cannot reliably engage with core mechanics, neither performance nor difficulty level informs their subjective experience.
\end{itemize}

\section{Discussion}
The \textit{Atari Games Challenge} experiment has shown considerable potential in laying the foundations for an extensive analysis of PX. Although limited to $19$ participants, the pilot study has allowed researchers to gather and triangulate data from multiple sources, which revealed interesting preliminary results from the exploratory analysis.

The models fitted through the dataset showed promising results in the explanation of game difficulty: similarities to flow were found, but only in performative challenge, finding much different relations, or even a lack of them, in other cases. This further shows that difficulty cannot be simply treated as an objective function, and should be analysed through multiple perspectives in order to make better DDA algorithms. Beyond that, the inclusion of OGD and SGD into a single analysis can potentially improve the balance of a game or the learning of mechanics. The dataset supported the finding of insights that go beyond ``optimising for engagement''. Further supported by the C-RTA, there are interesting bases for more promising insights about the nature of difficulty and how to model it.


By addressing a few remaining issues, such as the use of a more specialised survey for more detailed information on perceived challenge and the use of less abstract games, we intend to further enhance the experiment to strengthen the robustness and scale of the produced datasets.

Furthermore, the study presented in this paper focused primarily on testing the validity of the experimental protocol and performing an initial behavioural analysis. As a next logical step, we expect to engage in the analysis of the relationship between behaviour and self-reported data with the collected psychophysiological signals (EEG and Gaze).
This future work will use time-locked averaging of EEG and eye-tracking signals aligned to C-RTA-reported confusion moments to test for consistent alpha/beta power changes or fixation instability patterns within and across participants. We will benchmark these against patterns in GAMEEMO and AMuCS datasets to determine if they reveal novel neurocognitive signatures of cognitive challenge or validate existing arousal/valence markers. These insights will guide protocol refinements, such as adaptive difficulty triggers based on real-time psychophysiology, for scaled future studies.

\subsection{Limitations and Future Directions}
This pilot study establishes a protocol that can be scaled into a full multimodal PX investigation. A larger, demographically diverse cohort, including older players, non-gamers, and children, would test whether the observed patterns generalise beyond university students and reveal how different populations experience difficulty.
Future studies should adapt trial and tutorial durations to game genre: cognitive challenges like Word Zapper may benefit from extended 5--10 minute tutorials and 4--5 minute trials, while performative challenges like Boxing may remain well-suited to two-minute sessions.
Performance metrics could be refined by normalising scores against enemies spawned or applying within-subject z-score standardisation, providing cleaner separation between difficulty effects and genuine performance variation.
Survey instruments such as the CORGIS~\cite{denisova_measuring_2020} or NASA-TLX~\cite{hart_task_1988} would capture nuances the miniPXI missed, distinguishing frustration from focused engagement and attributing difficulty to skill deficits versus game design.
Hierarchical Bayesian models with participant-level random effects would properly account for the repeated-measures structure of multimodal data. Finally, structured C-RTA annotation pipelines and anonymisation techniques could enable public sharing of qualitative data alongside the quantitative dataset, enriching future multimodal PX research.

\section{Conclusion}
We have presented an experiment that answers the call for the need of more comprehensive datasets in PX research. \textit{Atari Games Challenge} allows the collection of game telemetry, questionnaires, psychophysiological, and qualitative data sources.
The conducted pilot study has highlighted the power of a synchronised, multimodal dataset in enabling a deep and comprehensive understanding of PX: it has shown how a multisourced analysis can broaden perspectives on a dimension such as difficulty, often considered as the mere performance of a player; and it has allowed us to identify confusion and its behavioural patterns during gameplay, with promising preliminary results.



\textit{Atari Games Challenge} represents another step towards an effective collection of multimodal data for the analysis of player experience. However, if we want to achieve a comprehensive understanding of it, it is necessary that more studies embrace this complex dimension.

\bibliographystyle{IEEEtranS}
\bibliography{references,references_pabu}

\end{document}